# Surface Orientation Strengthening in Metallic Nanowires

Haojie Wei [1, a)], Yueguang Wei [2]

[1] *Department of Engineering Mechanics, Henan University of Science and Technology, Luoyang 471023, China*

[2] *State Key Laboratory of Nonlinear Mechanics, Institute of Mechanics, Chinese Academy of Sciences, Beijing 100190, China*

a) Electronic mail: haojie.wei@hotmail.com

## ABSTRACT

Due to the large free surface and strong anisotropy, the mechanical property of nanowire differs from its bulk counterpart. In this paper, the strong dependence of NW's strength on its surface orientation is revealed, and two modes of dislocation nucleation have been identified: the corner nucleation and the surface nucleation. The activation energy and activation volume of surface nucleation are much larger than those of the corner nucleation. The intersection line between lateral surface and {111}-type slip plane causes the discrepancy of the nucleation modes for the first dislocation.

## 1. Introduction

Pure polycrystalline materials can be strengthened by either decreasing grain size[1,2] or introducing microstructures, such as coherent twin boundaries,[3] into the interior of grains. Such strengthening strategies have been proved effectively in nanowires (NWs). For example, NW can be strengthened by either decreasing its diameter[4,5] or introducing twin boundaries[6].

However, due to the effects of large free surfaces and strong anisotropy, the mechanical properties of NWs can be quite different from their bulk counterparts. Also, there may be some potential strengthening methods undiscovered in NWs, originating from such remarkable differences.

Many studies have revealed the effect of axial orientation on the mechanical properties of NWs. Deng has studied the mixed dependence of the strength on both the internal and external microstructures in metallic NWs[7]. However, the systematic study has not been done about the dependence of the strength of NWs on the lattice orientation of lateral surface.

In this paper, we address the effects of surface orientation on the strength of Cu NWs with [100] axial orientation and nearly square cross section using molecular dynamics simulations. We revealed the strong dependence of tensile strength of NWs on the lattice orientation of lateral surface. We proposed a new strengthening method in NWs, which can be called the surface orientation strengthening.

## 2. Simulation methods

Molecular Dynamics simulations are employed to simulate four Cu NWs. The NWs in our studies are all [100] oriented in the axial direction, with nearly square cross-section. The side length of the square cross-section and the axial length are 14a and 64a, respectively (a is the lattice constant of Cu). The only differences among these NWs are the lattice orientation of lateral surfaces. Thus the probabilities of size effect are excluded. To make the contrast more striking, the lateral surfaces of each NW are constructed in lattice planes of single type.

In fcc structures, {111}, {100}, and {110} are the first three closest-packed lattice planes, and previous experimental investigations have confirmed that {111} and {100} are the most commonly observed lattice planes in NWs. Therefore, {111} and {100} lattice planes are selected. Furthermore, in order to make the study more comparative, {110} and {102} lattice planes are also simulated. (In this paper, NWs taking {100}, {102}, {110}, and {111} lattice planes as their lateral surface are denoted by {100}-surface NW, {102}-surface NW, {110}-surface NW, and {111}-surface NW, respectively.)

The simulation models of {100}-, {102}- and {110}-surface NWs can be easily constructed by cutting from a single crystal Cu along [100] direction with different angles (See Fig. 1). {111}-surface NW is curved out on the basis of {110}-surface NW by removing atoms beyond the {111} planes intersecting the lateral surface, similar to the surface morphology of Au NW used in ref.[8]

The embedded atom method (EAM) potential for Cu developed by Mishin[9] was adopted, for its accurate prediction of the surface energy and dislocation nucleation. All simulations were performed at 300K using a Nose–Hoover thermostat. Periodic boundary condition was imposed along the axial direction, while the lateral surfaces were kept free during the whole simulation process for all NWs.

Prior to deformation, each model was relaxed under zero pressure in the axial direction, governed by NPT ensemble. After obtaining equilibrium configuration, uniaxial tension was applied along the axial direction until the strain is up to 0.16. The strain rate was kept at a constant of $2.0 \times 10^8$ s$^{-1}$, and NVT ensemble governed the whole loading process. Common neighbor analysis was used to detect the dislocation from perfect fcc lattice structures.

## 3. Results

Tensile stress-strain curves of NWs simulated are shown in Fig 2. The stress is averaged within 1 picosecond (ps) to eliminate the interference of thermal fluctuations.

The stress-strain curve exhibited a marked elastic section and a steep yield point. In the initial stage of loading, the stress increases linearly with the increase of strain. At the end of elastic stage, a sharp yield happened and the stress dropped dramatically. Therefore, the yield point coincides with the peak point, and both yield stress and maximum stress are identical. The yielding of Cu NW is associated with the nucleation of (1/6) [112] leading partial from surface. This is consistent well with the previous MD simulations of fcc nanowires under uniaxial tension.

The yielding stress of {102}-surface NW, {100}-surface NW, {110}-surface NW, and {111}-surface NW are 5.77, 7.24, 7.36, and 8.42 GPa, respectively. The remarkable difference in strength indicates the strong dependence of strength on the surface orientation.

The ideal strength of Cu single crystal in [100] orientation has also been calculated for comparison, which is 9.07GPa. The yield strength of {111}-suface Zigzag NW is near the ideal strength. Thus, near ideal strength in [100] orientation can be obtained by only designing the surface microstructure of NW.

## 4. Discussion

The strength difference among these NWs can be naturally attributed to the surface energy density, originated from the surface orientation. Lowest energy principle is a general rule of nature. A correspondence has been established between the "surface stability" and "surface energy", i.e. the lower energy the more stable. Thus NW with lower surface energy density in lateral surface will be more stable, the first dislocation will nucleate more difficultly, and consequently the NW will possess higher strength.

In FCC metals, the surface energy density decreases in the order: {102} > {110} > {100} > {111}. However, the strength increases in the order: {102}-surface NW < {100}-surface NW < {110}-surface NW < {111}-surface NW, which is not entirely coincident with the sequence of surface energy density.

Therefore, there should be some other reasons besides the surface energy density to cause the obvious difference in tensile strength. By careful analyzing the dislocation nucleation process, we found out the reason.

The nucleation sites of the first dislocation in different NWs simulated are shown in the upper image of Fig. 3. From Fig. 3, we can easily find the difference. In {102}-surface NW and {100}-surface NW, the first dislocation nucleates from the sharp corner of one of the four equivalent edges; while in {110}-surface NW, it nucleates from the side surface; and in Zigzag {111}-surface NW, the first dislocation nucleates from the concave edge where two {111} facets intersects. In terms of dislocation nucleation region, the dislocation nucleates from a corner region in {102}-surface NW and {100}-surface NW, while it does from a belt region in {110}-surface NW and {111}-surface NW. To distinguish, the two dislocation nucleation modes can be called corner nucleation and belt nucleation, respectively.

Further analysis revealed the correspondence between the nucleation mode and the geometric morphologies of NW. Illustrated in the lower image of Fig. 3, the morphologies of intersection planes can be divided into two categories: Sharp corners generate when {100}-surface NW intersects with the {111} slip plane; while it forms two [110]-oriented lines perpendicular to the axis of NW when {110}-surface NW intersects with the slip plane. The two modes of dislocation nucleation are just coincident with the two morphologies of inclined cross section.

Dislocation nucleation is a stress-induced thermal activation process. Free-end string method[10,11] has been used to compute the minimum energy path (MEP) for dislocation nucleation. The zero-T MEP curve are shown in Fig. 4(a) and Fig.5(a) for dislocation nucleation in {100}-surface NW and {111}-surface NW, respectively. The saddle point structures shown in Fig. 4(d) and Fig. 5(e) coincide with our geometric analysis.

The activation energy and activation volume in belt nucleation is twice that of corner nucleation, which is coincident with the crack nucleation studies on semiconductor NWs.[12]

The zero-T activation energies computed at different stresses and the fitting curves by taking the functional form $E_b^0 = A\left[1 - e^{(1-\sigma/\sigma_{ath})}\right]$ are shown in Fig. 6(a). The activation volume as a function of stress calculated by differentiating the fitting curve in Fig. 6(a) are shown in Fig. 6(b). The values are the same order as the former studies on surface dislocation nucleation.[10,13]

To consider the temperature effect on the activation energy, we take the following approximation: $E_b = (1-\frac{T}{T_m})E_b^0$, where $T_m$ is the surface disordering temperature. Also, we take $\Omega = (1-\frac{T}{T_m})\Omega^0$ for considering the temperature effect on the activation volume.

Numerical solving equation $\frac{E_b(\sigma,T)}{k_B T} = \ln\frac{k_B T N v_0}{E\dot\varepsilon\Omega(\sigma,T)}$ by taking different strain rate and temperature, the dependence of tensile strength on strain rate and temperature can be obtained.

Different lattice orientation of lateral surface leads to different intersection plane between lateral surface and {111} slip planes, and consequently leads to the difference of dislocation nucleation modes. Different nucleation modes need different activation energies, which finally determine the yield strength of NWs.

The higher the activation energy is, the higher strength the NW possesses. Then it becomes easy to explain the strength results. The strength of {100}-surface NW and {102}-surface NW are lower than that of {110}-surface NW and {111}-surface NW, for the dislocation nucleates from the corner and need lower activation energy in the former. As to {110}-surface NW and {111}-surface NW, the first dislocation both nucleate from the belt region. However, the potential sites for dislocation nucleation in Zigzag {111}-surface NW is far less than that of {110}-surface NW, for the dislocation can only nucleate from the concave intersection line of two {111} planes in the former. Therefore, Zigzag {111}-surface NW possesses the highest strength.

## 5. Summary

In summary, we have revealed the strong dependence of the strength of [100] oriented NWs on lattice orientation of lateral surface by MD simulations. The strength of NW is not only determined by the surface energy density, but also by the nucleation mode of the first dislocation.

Due to the difference of surface orientation, the intersection line between lateral surface and {111} slip plane exhibit different morphologies. Such morphology difference causes the different mode of dislocation nucleation, which furthermore leads to the difference of the activation energy required for dislocation nucleation.

Based on the simulations, we propose a new strengthening mechanism in NWs, the surface orientation strengthening, which is originated from the large free surface and the strong anisotropy. Near ideal strength in [100] orientation can be obtained only through the microstructure design of surface orientation.

## Acknowledgements

The computational resources were provided by Supercomputing Centre of Chinese Academy of Science (SCCAS). All simulations were performed using Lammps Molecular Dynamics Simulator[14] and visualized by Atomeye[15].

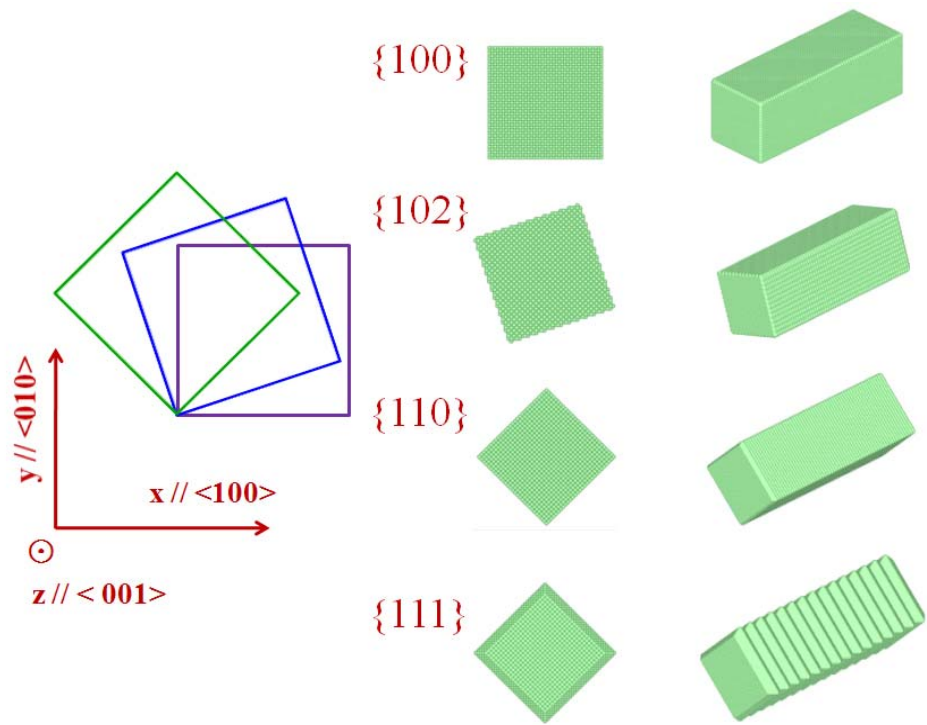

Fig. 1. The four Nanowires studied in this paper. All of them are <001> oriented in the axial direction, while the lattice orientation of lateral surfaces are different.

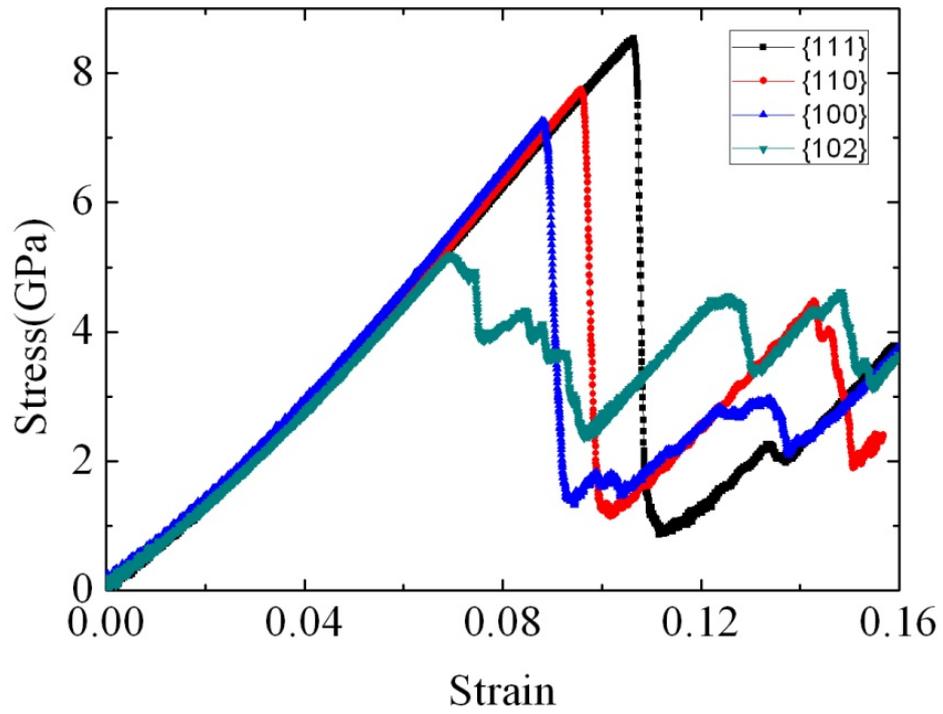

Fig. 2. The stress-strain curves of unaxial tension. The sharp yield is associated with the nucleation of (1/6) [112] leading partial from surface.

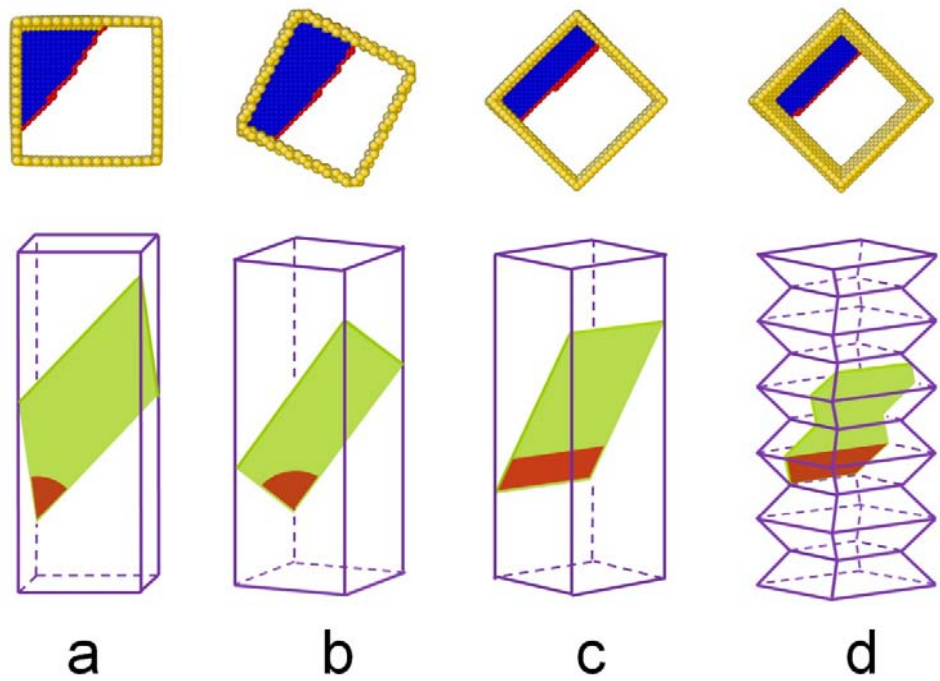

Fig. 3. The atomic configuration after the initial dislocation nucleation (upper image: yellow for surface atoms, red for dislocation core atoms, and blue for stacking fault atoms) and geometric morphology of the intersection plane between the {111}-slip plane and the surface of nanowires (lower image): (a) {100}-surface NW, (b) {102}-surface NW, (c) {110}-surface NW, and (d) {111}-surface NW.

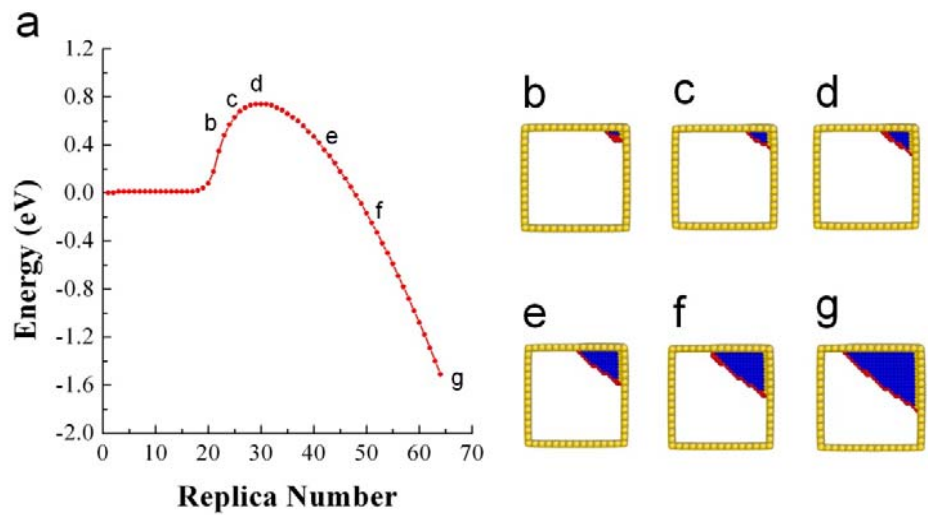

Fig. 4. Minimum Energy Path (MEP) calculation of {100}-surface NW with σ = 4.79GPa. (a) The energy curve as a function of replica number. (b)-(g) Different states along MEP, corresponding to the energies in (a). (d) is the saddle point of MEP, shown corner nucleation mode.

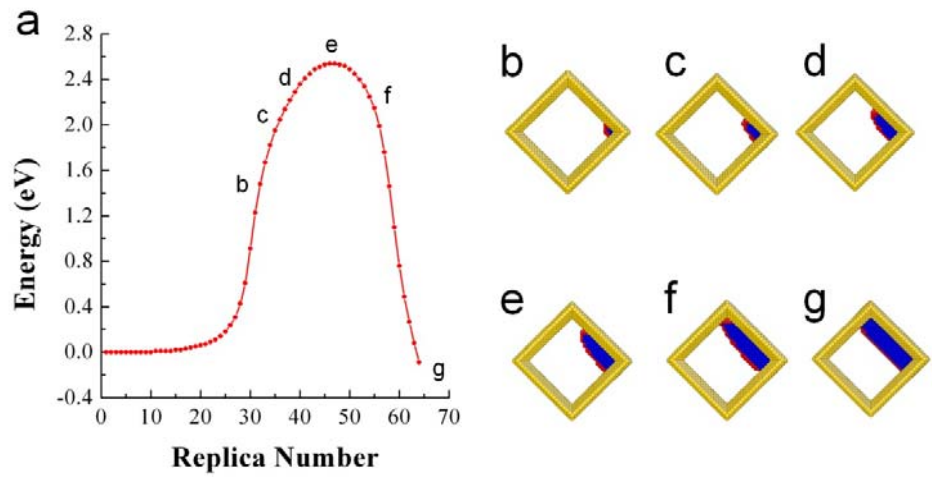

Fig. 5. Minimum Energy Path (MEP) calculation of {111}-surface NW with σ = 4.45GPa. (a) The energy curve as a function of replica number. (b)-(g) Different states along MEP, corresponding to the energies in (a). (e) is the saddle point of MEP, shown belt nucleation mode.

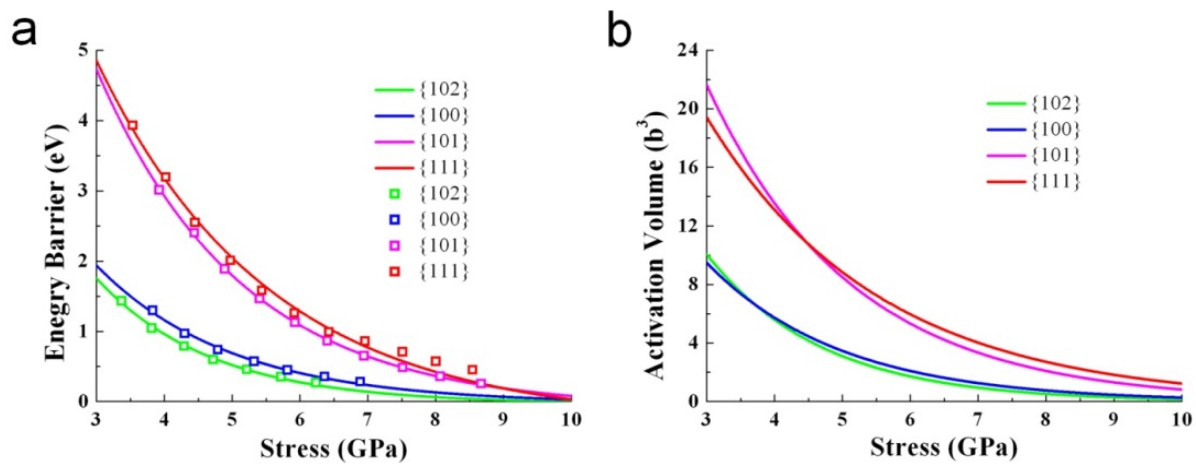

Fig. 6. Stress-dependence activation parameters for (a) Energy barrier and (b) Activation Volume. The points in (a) are computed by Energy Barrier calculations and the solid lines are the fitting curve.